\documentclass[aps,showpacs,twocolumn]{revtex4}
\usepackage{amsfonts}
\usepackage{amsmath}
\usepackage{amssymb}
\usepackage{graphicx}
\usepackage{bm}
\usepackage[export]{adjustbox}

\input{tcilatex}

\begin{document}

\title{Self-diffusiophoresis of Janus particles in near-critical mixtures}
\author{Alois W\"{u}rger}
\affiliation{Laboratoire Ondes et Mati\`{e}re d'Aquitaine, Universit\'{e} de Bordeaux \&
CNRS, 351 cours de la Lib\'{e}ration, 33405 Talence, France}

\begin{abstract}
We theoretically study the self-propulsion of a laser-heated Janus particle
in a near critical water-lutidine mixture, and relate its velocity $v_{p}$
and squirmer parameter $\beta $ to the wetting properties of its two
hemispheres. For non-ionic surface forces, the particle moves the active cap
at the front, whereas a charged hydrophilic cap leads to backward motion, in
agreement with experiment. Both $v_{p}$ and $\beta $ show non-monotonic
dependencies on the heating power, and may even change sign. The variation
of $\beta$ is expected to strongly affect the collective behavior of dense
squirmer systems.

PACS\ numbers 05.70.Ln; 66.10.C-; 82.70.-Dd
\end{abstract}

\maketitle

In recent years, artificial microswimmers have been realized by Janus
particles which move along the concentration or temperature gradients
generated by their own chemical or thermal activity \cite%
{Pax05,How07,Jia10,Vol11,Bar13}. Oriented autonomous motion has been
achieved throug dynamical feedback \cite{Qia13} or rectification in a
periodically structured channel \cite{Gho13}, opening applications such as
targeted transport and pumping of passive particles. In dense systems,
active Janus particles aggregate in dynamical clusters \cite{The12,But13}.
This observation was related to short-range hydrodynamic effects \cite%
{Ish08,Llo10,Zoe14}, in terms of the squirmer model originally developed for
the motility of bacteria. Self-propulsion mechanisms have generally a strong
diffusiophoretic component \cite{Ebb10,Kap13}; in the case of ionic
molecular solutes, self-generated electric fields and ion effects may
contribute to the motion \cite{Mor10,Bro14}.

Diffusiophoresis was first rationalized by Derjaguin et al. \cite{Der47},
when observing that wax particles dispersed in a non-uniform glucose
solution, migrate toward lower sugar concentration. Because of its
unfavorable interaction with wax ($u>0$), sugar is depleted in the boundary
layer, the adsorption parameter\ $\Gamma =\int_{0}^{\infty
}dzz(e^{-u/k_{B}T}-1)$ is negative, and the wax particle migrates to lower
glucose content at the velocity \cite{Der87} 
\begin{equation}
v_{p}=\frac{2k_{B}T}{3\eta }\Gamma \nabla n,  \label{2}
\end{equation}%
where $\eta $ is \ the solvent viscosity. For a molecular solute that is
attracted by the surface, one has $\Gamma >0$ and the particle moves toward
higher concentration.

There is no such simple rule for self-propelling Janus particles, where both
the adsorption parameter and the concentration gradient vary along the
particle surface, and where catalytic coating may result in a multicomponent
boundary layer \cite{Ebb10,deG14}.\ A particularly intricate situation
occurs for hot Janus particles in a near-critical water-lutidine mixture 
\cite{Vol11}, which migrate in a self-generated composition gradient. Though
their motion is clearly related to the wetting properties of their active
and passive surfaces \cite{But12,But13}, there is at present no explanation
for the sign and magnitude of the velocity.

\begin{figure}[ptb]
\includegraphics[width=0.4\columnwidth,valign=t]{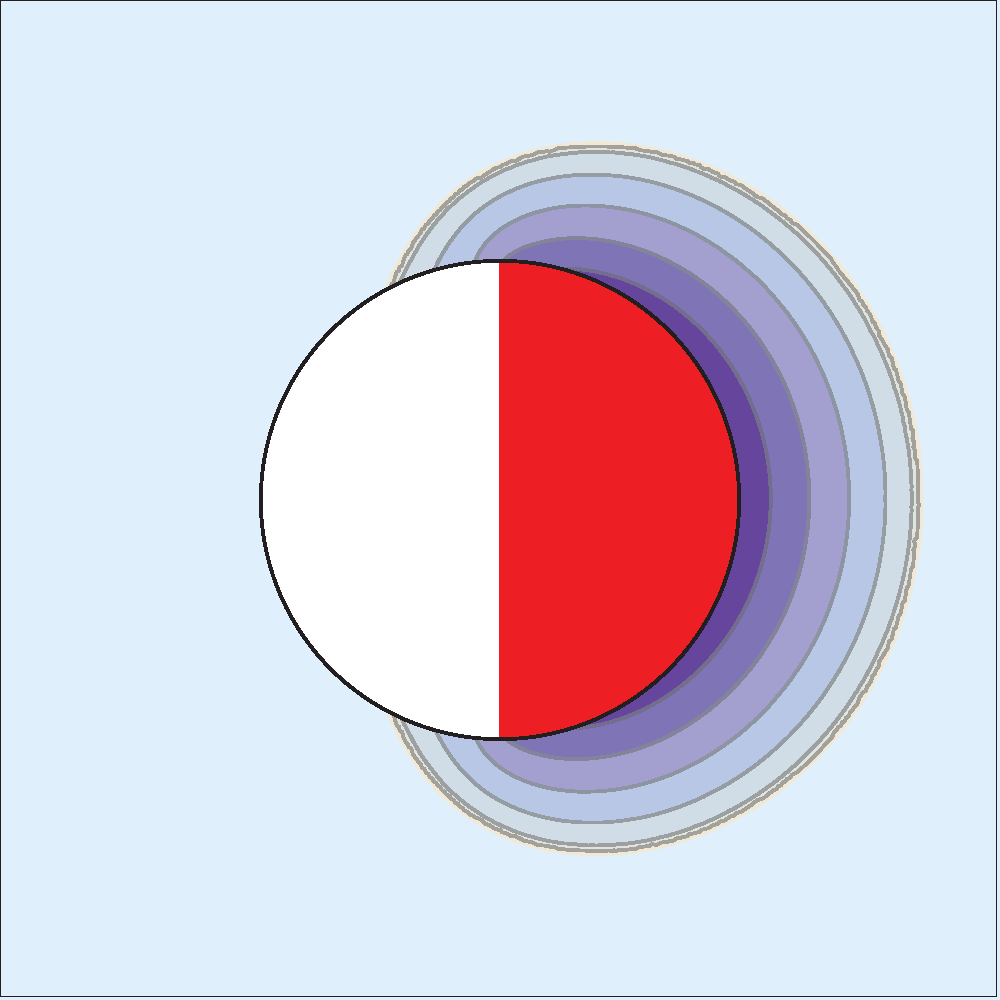} %
\includegraphics[width=0.56\columnwidth,valign=t]{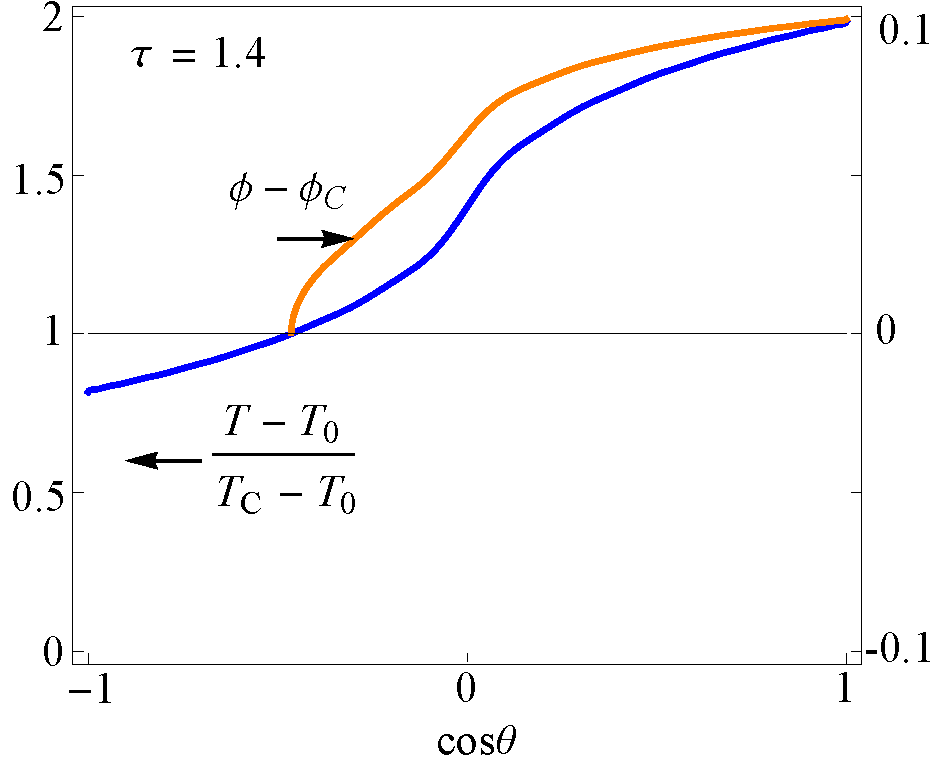}
\caption{Critical droplet ($\protect\phi >\protect\phi _{C}$) surrounding a
particle with hydrophilic surface coating on both the cap (red) and the
uncapped hemisphere (white). The surface temperature profile $T(\mathbf{r)}$
(calculated from the thin-cap limit of Ref. \protect\cite{Bic13}) and the
composition (from (\protect\ref{4}) with $C=100$ ${}^\circ$C) are plotted as
a function of the cosine of the polar angle. The scale of the reduced
temperature $(T-T_0)/(T_C-T_0)$ is on the left ordinate, and that of the
water content $\protect\phi-\protect\phi_{C}$ on the right one. The
parameter $\protect\tau=(T_m-T_0)/(T_C-T_0)$ depends on the temperature at
midplane $T_m$.}
\end{figure}

In this Letter we study self-diffusiophoresis in such near-critical binary
mixtures \cite{Vol11,But12,But13}. Starting from the properties of the
demixing volume surrounding the particle, we evaluate the velocity $v_{p}$
and the squirmer parameter $\beta $, both of which depend in an intricate
manner on the heating power and the dispersion forces exerted by the two
hemispheres. Finally, we compare with recent experiments and discuss charge
effects.

\textit{The critical droplet. \ }Fig. 1 illustrates an active Janus particle
in a water-lutidine mixture at the critical water content $\phi _{C}=0.72$
and at a bulk temperature $T_{0}$ which is slightly below the critical value 
$T_{C}=34.1$ ${}^\circ$C. Illuminating the particle with a laser beam,
results in a temperature profile $T(\mathbf{r)}$ that exceeds $T_{C}$ on
part or all of the surface, and thus causes local demixing. Assuming a
quadratic relation to the local composition, $T-T_{C}=C(\phi -\phi _{C})^{2}$%
, one finds the change of water content 
\begin{equation}
\phi (\mathbf{r})-\phi _{C}=\pm \sqrt{\frac{T(\mathbf{r})-T_{C}}{C}},
\label{4}
\end{equation}%
where the two signs correspond to water-rich and lutidine-rich phases, and
where $C\sim 100$ $%
{{}^\circ}%
$C \cite{Gra93}. This mean-field relation ceases to be valid at the critical
point where composition fluctuations become long-range.

\begin{figure}[ptb]
a) \includegraphics[width=0.38\columnwidth,valign=t]{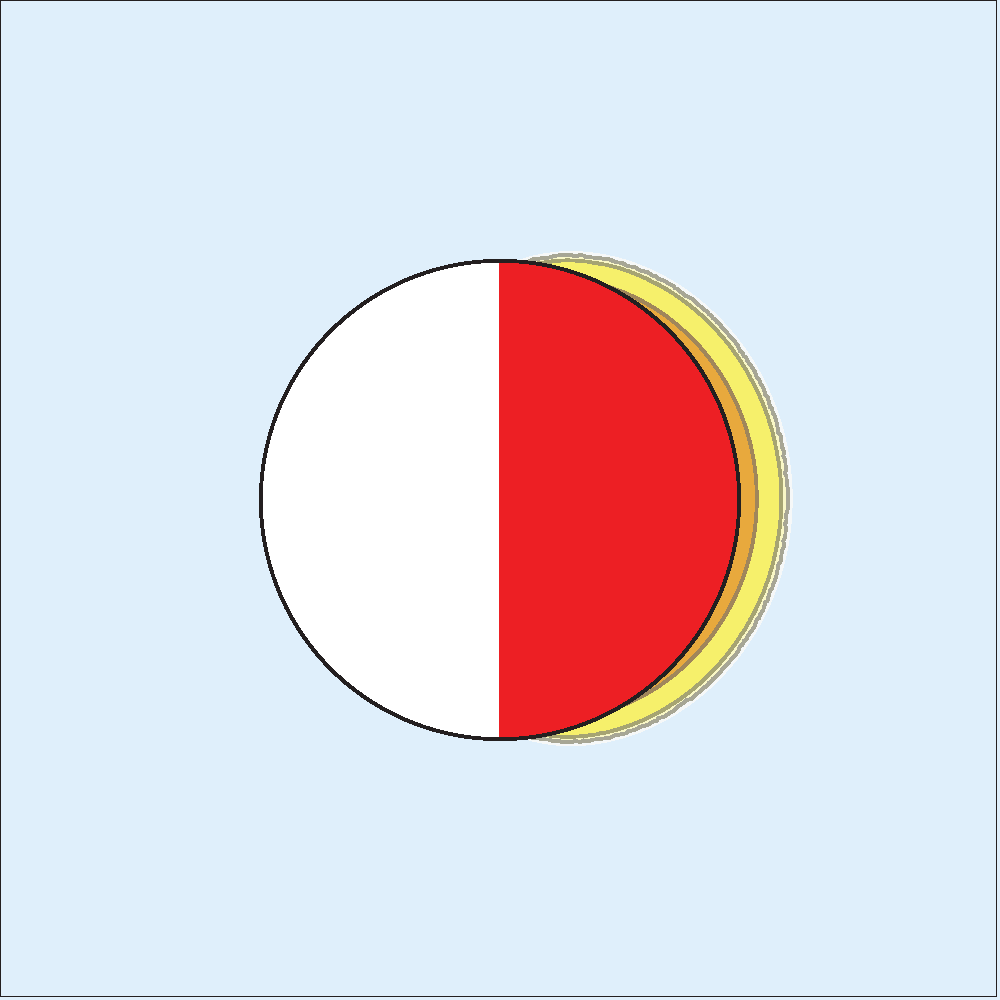} %
\includegraphics[width=0.54\columnwidth,valign=t]{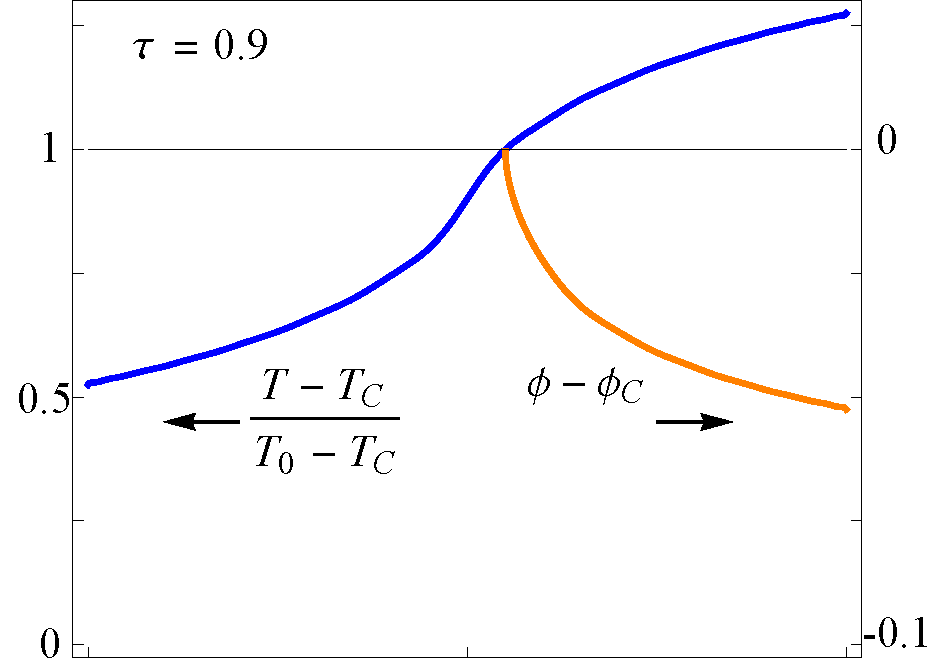} b) %
\includegraphics[width=0.38\columnwidth,valign=t]{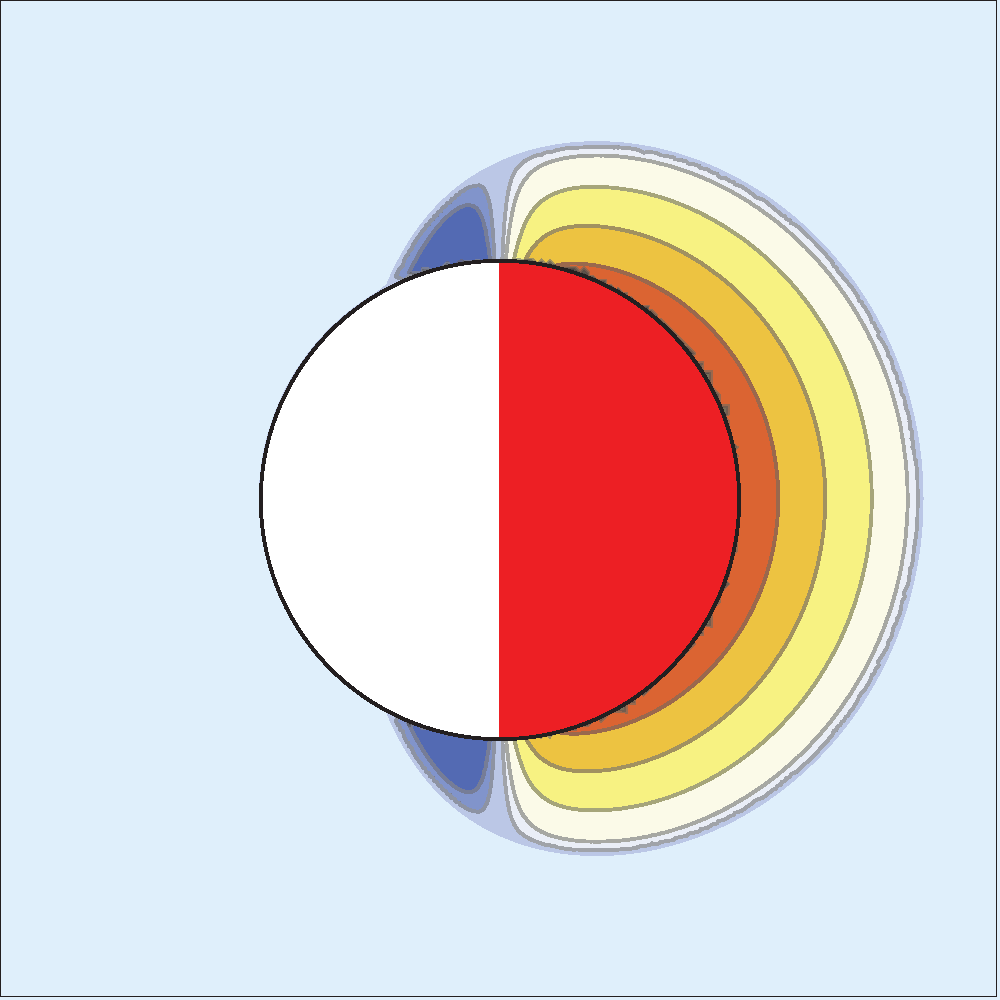} %
\includegraphics[width=0.54\columnwidth,valign=t]{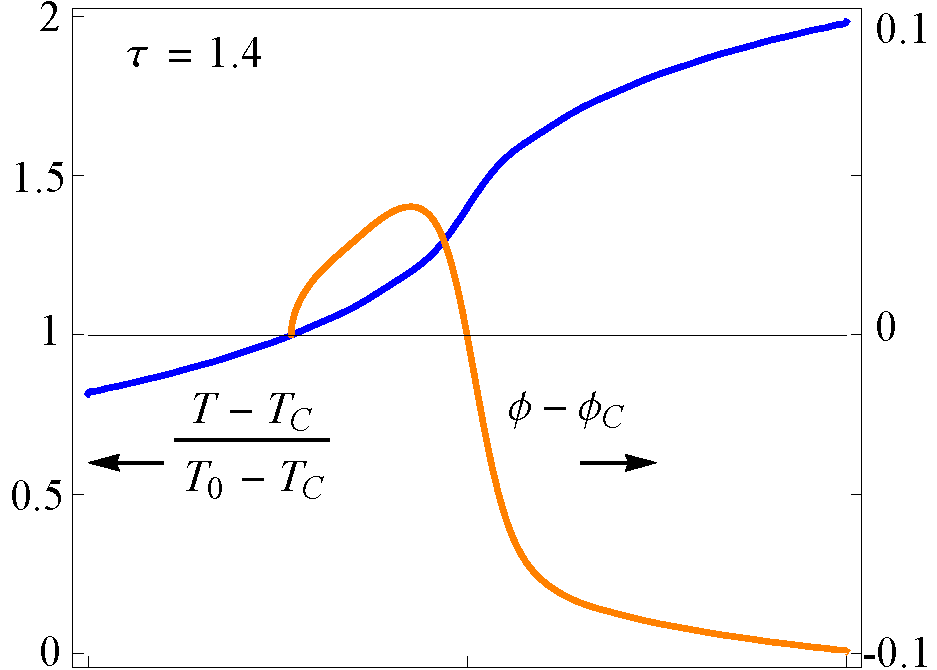} c) %
\includegraphics[width=0.38\columnwidth,valign=t]{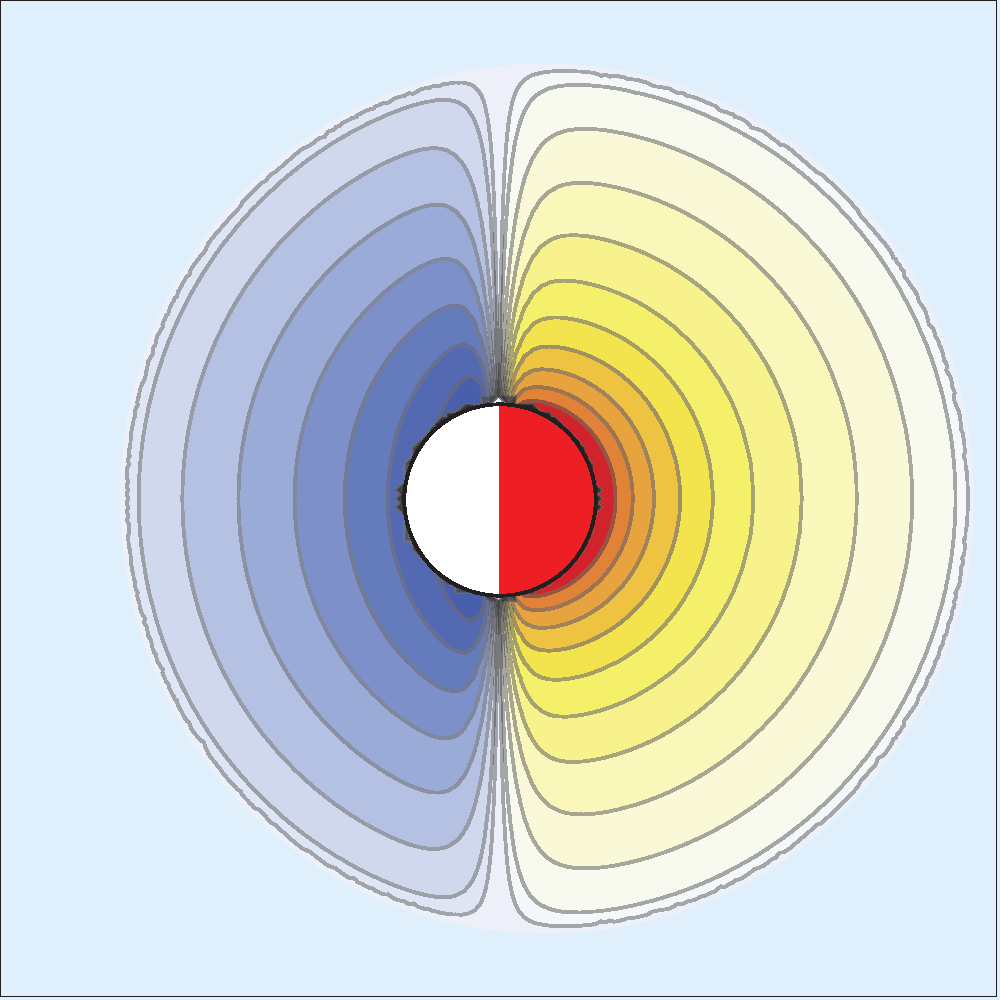} %
\includegraphics[width=0.54\columnwidth,valign=t]{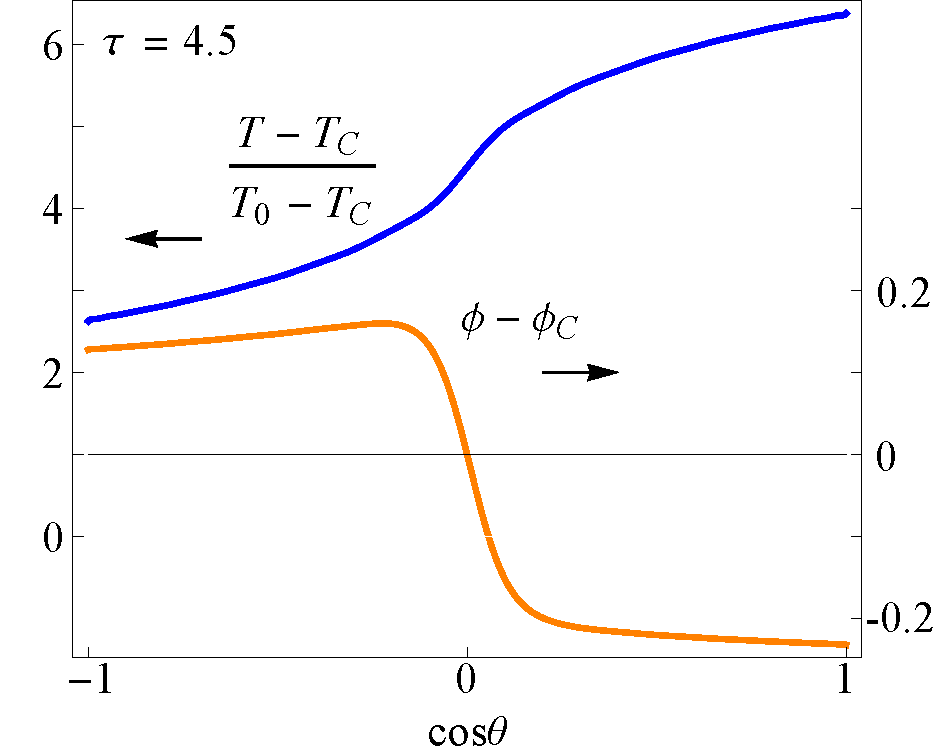}
\caption{Demixing volume surrounding a particle with hydrophobic cap and
hydrophilic remainder. Heating power increases from a) to c); the right
panels show temperature and composition profiles as in Fig. 1. a) The
critical condition is satisfied only in a thin lutidine-riche droplet that
partly covers the cap. b) Most of the surface is above $T_C$; the cap is
surrounded by the lutidince-rich phase, and the hydrophilic remainder by the
water-rich phase. The phase boundary is modelled by a factor $\tanh(\cos%
\protect\theta/c_0)$ with $c_0=0.1$. c) Strong heating leads to $T>T_C$ on
the entire surface and to a spherical demixing volume. }
\end{figure}

A more complex situation occurs for a particle with a hydrophobic cap, as
illustrated in Fig. 2. If $T_C$ is reached on both hemispheres, the critical
volume splits in lutidine-rich and water-rich compartments, with the phase
boundary attached at the particle's midplane \cite{Yu14}. This separation is
possible since the interface tension between the spinodal phases, $\gamma
<10^{-4}$ N/m for $T<36$ ${}^\circ$C \cite{Gra93}, is much smaller than the
particle's surface energy. For the strong-heating case (c), the critical
volume is almost spherical, and its radius $\tau a$ much larger than that of
the particle, $a$. Experiments cover the whole range from a small demixing
area to a large critical droplet of tens of microns \cite{But12}.

\textit{Slip velocity. }Contrary to the demixing volume, the boundary layer
is not in a quiescent state, but shows a non-uniform pressure and a steady
diffusion current. Assuming that mutual diffusion of water-lutidine is faster 
than advection and inserting the current in Stokes' equation, one finds the
effective slip velocity \cite{supplmat},%
\begin{equation}
v_{s}=-\frac{k_{B}T}{\bar{v}\eta }\Gamma \frac{d\phi }{dx},  \label{24}
\end{equation}%
where we have defined the mean inverse molecular volume $\bar{v}^{-1}=\phi
_{C}v_{l}^{-1}+(1-\phi_{C})v_{w}^{-1}$ and the adsorption parameter 
\begin{equation}
\Gamma =\int_{0}^{\infty }dzz\frac{e^{-\psi _{w}}-e^{-\psi _{l}}}{\phi
e^{-\psi _{w}}+(1-\phi )e^{-\psi _{l}}}.  \label{25}
\end{equation}%
The effective potential of water and lutidine, $\psi _w$ and $\psi _l$ are
given in units of the thermal energy, and vanish well beyond the interaction
range $\lambda $. In the dilute limit $\phi \rightarrow 0$ and with $%
u=k_{B}T(\psi _{w}-\psi _{l})$, one recovers Derjaguin's adsorption factor
in (\ref{2}). Note that that the slip velocity $v_{s}$ does not depend on
the parallel force component $\partial _{x} \psi$, but on the composition
gradient only \cite{Wue10}. 

For an order-of-magnitude estimate, it is convenient to explicit the
adsorption parameter for a square well potential of width $\lambda $ and
prefactor $\bar{\psi}$, where the integral in (\ref{25}) gives $\frac{1}{2}%
\lambda ^{2}$ with $\psi _{i}\rightarrow \bar{\psi}_{i}$ in the second
factor. A strongly hydrophobic surface repels water and attracts lutidine ($%
\bar{\psi}_{w}>\bar{\psi}_{l}$), such that $\Gamma <0$, whereas a
hydrophilic surface is characterized by $\Gamma >0$. With typical parameters 
\cite{But12} we find $\Gamma \sim 10^{-21}$ m$^{2}$ and, supposing an
interaction length $\lambda $ of a few \AA , we deduce $\bar{\psi}_{i}\sim
10^{-2}$.\ Then the adsorption parameter $\Gamma =\frac{1}{2}\lambda ^{2}(%
\bar{\psi}_{l}-\bar{\psi}_{w})$ takes a constant value $\Gamma _{\text{cap}}$
on the cap, and $\Gamma _{\text{unc}}$ on the remainder.

\textit{Self-propulsion.} \ The particle velocity is obtained by averaging
the slip velocity over the surface, $\mathbf{v}_{p}=-\left\langle \mathbf{v}%
_{s}\right\rangle $ \cite{And89}. For an axisymmetric particle one finds 
\begin{equation}
v_{p}=\frac{k_{B}T}{2\bar{v}\eta a}\int_{-1}^{1}dc(1-c^{2})\Gamma \partial
_{c}\phi ,  \label{30}
\end{equation}%
where we have used the relation $\partial _{x}=a^{-1}\sqrt{1-c^{2}}\partial
_{c}$ between the local coordinate $x$ and the cosine of the polar angle $%
c=\cos \theta $. Sign and magnitude of the velocity are determined by the
product of the adsorption factor $\Gamma $ and the derivative of the
composition gradient $\partial _{c}\phi $. A particularly complex behavior
occurs for cases as in Fig. 2, where $\Gamma _{\text{cap}}$ and $\Gamma _{%
\text{unc}}$ take opposite signs and where the gradient $\partial_{c}\phi $
is largest in the midplane area.

\begin{figure}[ptb]
\includegraphics[width=\columnwidth]{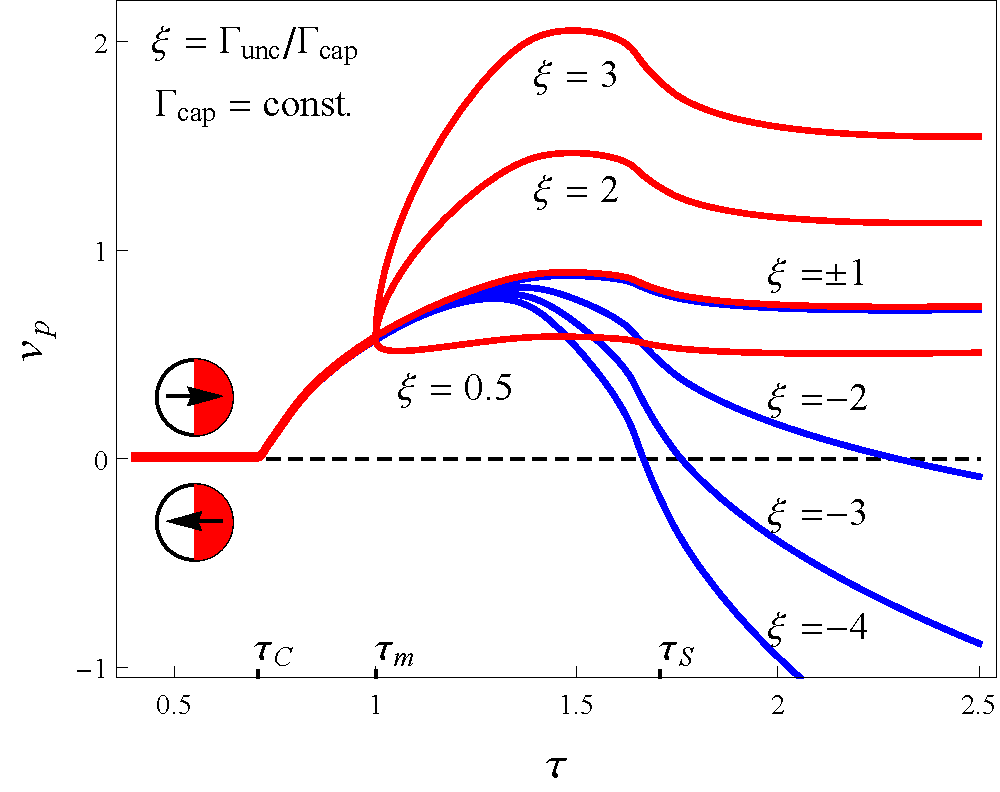}
\caption{Self-propulsion velocity $v_{P}$ as a function of $\protect\tau %
=(T_{m}-T_{0})/(T_{C}-T_{0})$, for different values of absorption
parameters. Positive $v_{p}$ means that the particle moves forward (cap at
the front). We fix the adsorption parameter of the cap and vary the ratio $%
\protect\xi =\Gamma _{\text{unc}}/\Gamma _{\text{cap}}$. Self-propulsion
sets in at $\protect\tau _{C}=1/\protect\sqrt{2}$. If the critical droplet
extends to both hemispheres ($\protect\tau >1$), the velocity strongly
depends on $\protect\xi $ and changes sign for $\protect\xi <-1$. The velocity scale is of the order of $\mu$m/s.}
\end{figure}

In Fig. 3 we plot the particle velocity as a function of the surface
temperature at midplane $T_{m}$, in terms of the reduced quantity $%
\tau=(T_{m}-T_{0})/(T_{C}-T_{0})$. Critical conditions on the summit of the
cap are reached at $\tau_{C}=1/\sqrt{2}$, on the particle's midplane at $%
\tau _{m}=1$, and on the entire surface at $\tau _{S}=1+1/\sqrt{2}$. The
behavior of the velocity is to a large extent determnined by the ratio of
adsorption parameters $\xi =\Gamma _{\text{unc}}/\Gamma _{\text{cap}}$; that
is, by the wetting properties of the two hemispheres. We distinguish three
parameter ranges.

(i) For $\tau _{C}<\tau <1$, the critical droplet covers only part of the
cap, as in Fig. 2a. Since in this range, $\Gamma $ and $\partial _{c}\phi $
carry the same sign, the particle moves forward ($v_{p}>0$) for both
hydrophobic and hydrophilic coating. The velocity is independent of $\Gamma
_{\text{unc}}$.

(ii)\ In the range $1<\tau $ and $\xi >0$, both hemispheres contribute to
the integral in (\ref{30}). The cusp at $\tau _{m}$ occurs because of the
large derivative $\partial _{c}\phi $ on the uncapped hemisphere; see the
right panel of Fig. 1. The velocity goes through a maximum at $\tau \approx
1.5$, where most of the particle is covered by the critical volume; at
strong driving $\tau\gg1$, it increases as $v_{p}\propto(1+\xi)\sqrt{\tau }$ 
\cite{supplmat}.

(iii) For $1<\tau $ and $\xi <0$, the velocity is to a large extent
determined by the change of sign of $\phi -\phi _{C}$ at midplane, where the
contributions of the two hemispheres partly cancel in (\ref{30}).\ If the
adsorption parameter is larger on the uncapped part, $\xi <-1$, it dominates
the velocity and finally results in a change of sign; well beyond $\tau _{S}$
one finds $v_{p}\propto (1-\xi )\sqrt{\tau }$ \cite{supplmat}.

Several features can be traced back to the relation between the laser
intensity $I$ and the excess temperature. With $T_{m}-T_{0}=I\chi a/2\kappa $%
, the heat conductivity $\kappa $, and the absorption coefficient per unit
area $\chi $, one finds 
\begin{equation}
v_{p}\propto \frac{f(I-I_{C})}{\sqrt{a}},\ \ \ \ I_{C}=\frac{\sqrt{2}\kappa 
}{\chi a}\left( T_{C}-T_{0}\right).  \label{30a}
\end{equation}%
In a very narrow range above $\tau_C$ one has $f(x)=x^{3/2}$ (invisible in
Fig. 3); the cusp at $\tau_m$ follows the law $\sqrt{I-I_m}$.

We briefly discuss the above result in view of recent experiments. At small
or moderate driving we expect the particles to move the cap at the front.
This agrees with observations on carbon-capped silica beads \cite{But13} and
gold-capped particles with hydrophobic coating \cite{But12}. In addition,
the size dependencies of (\ref{30a}) and the overall shape of $v_{p}$ agree
well with the data of Buttinoni et al. (Fig. 4a of \cite{But12}), measured
for beads of different radius ($a=0.5\mu$m and $2.13\mu$m) in the range $%
\tau\leq1$ \cite{supplmat}. A strong discrepancy occurs for particles with hydrophilic caps,
which move the cap at the rear, at both weak \cite{Vol11} and strong driving 
\cite{But12}.

Some aspects of Fig. 3 disagree with a very recent theory paper \cite{Sam15}. As two main differences 
we note that Ref. \cite{Sam15} (i) assumes an isothermal cap with zero slip velocity and (ii) considers quite 
large velocities $v_p>10\mu$m/s, where the water-lutidine kinetics is governed by advection rather than diffusion \cite{supplmat}.

\begin{figure}[ptb]
\includegraphics[width=\columnwidth]{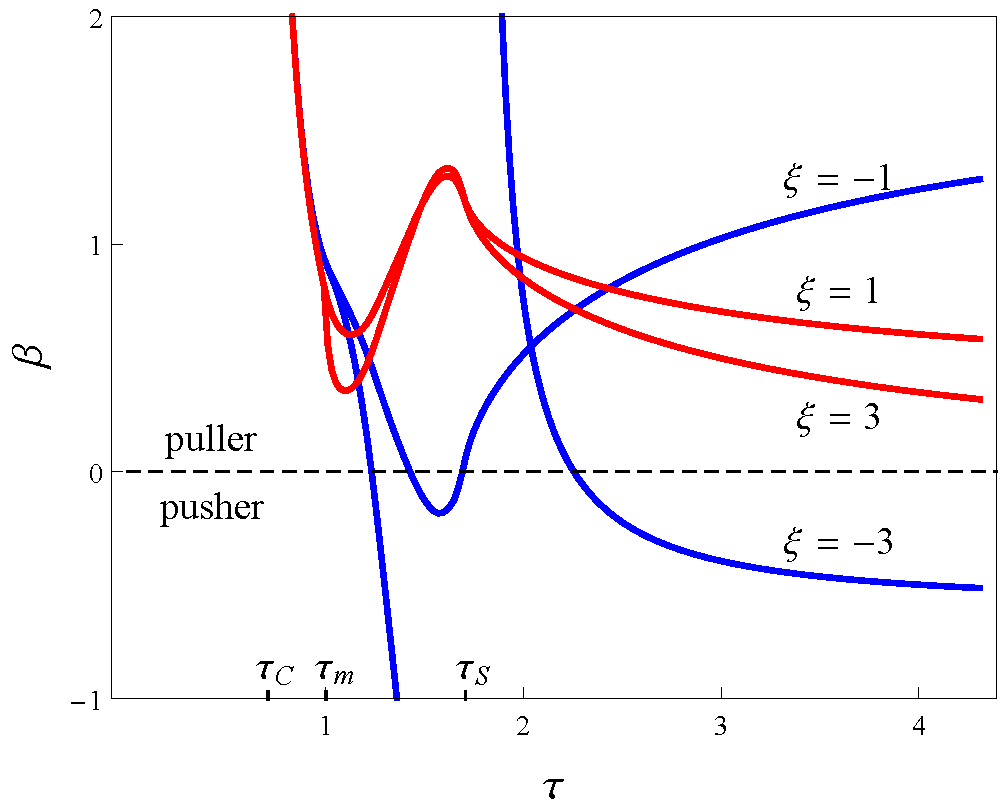}
\caption{Squirmer parameter $\protect\beta $. For $\protect\tau \rightarrow 
\protect\tau _{C}$, the active area reduces to a small spot, resulting in $%
\protect\beta =5$ \protect\cite{supplmat}. The divergency for $\protect\xi %
=-3$ occurs where the velocity $v_{p}$ is zero. }
\end{figure}

\textit{Squirmer parameter} The interaction of a microswimmer with a wall
and collective effects are to a large extent determined by the squirmer
parameter $\beta$ \cite{Ish08,Llo10,Zoe14}, which is defined through the
even component of the slip velocity $v_{s}=v_{s}^{0}\sin \theta (1+\beta
\cos \theta )$ \cite{Bla71}. A \textquotedblleft puller\textquotedblright\
is propelled by the activity of its front hemisphere ($\beta >0$), and a
\textquotedblleft pusher\textquotedblright\ by its back part ($\beta <0$).

In Fig. 4 we plot $\beta $ as a function of $\tau $ for different adsorption
parameter ratios $\xi $. At the onset of self-propulsion, where the active
area is reduced to a small spot at the summit of the cap, one finds%
\begin{equation}
\beta =5\ \ \ \ \ (\tau\rightarrow \tau_{C}).  \label{31}
\end{equation}%
(For a particle moving the active spot at the back, one has $\beta =-5.$)
With increasing driving, $\beta $ decreases rapidly and strongly depends on
the reduced temperature $\tau $ and the parameter $\xi $. Opposite
affinities of the two hemispheres, $\xi <0$, may result in pullers or
pushers of variable strength; the singularity for $\xi =-3$ occurs where the
particle velocity changes sign. This means that a tiny change in the driving
could significantly modify $\beta $ and thus the collective behavior \cite%
{Ish08,Llo10,Zoe14}.

\begin{figure}[ptb]
\includegraphics[width=\columnwidth]{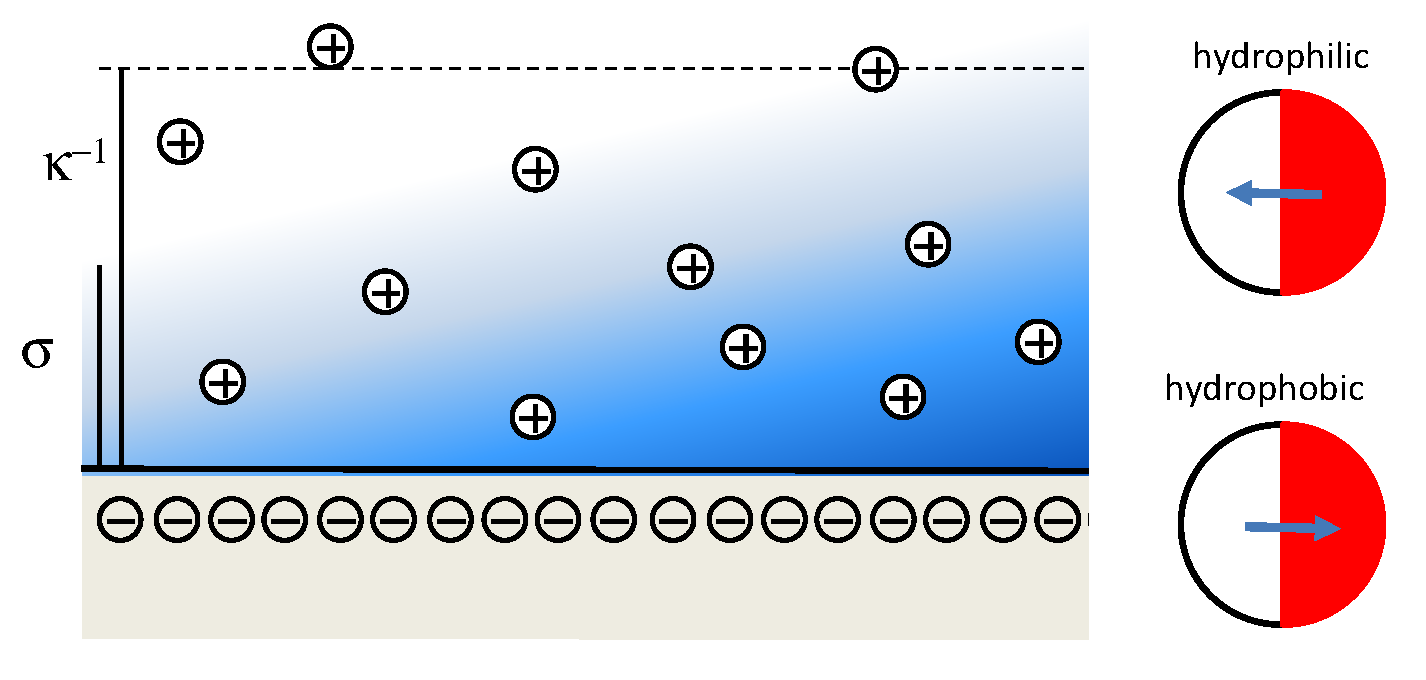}
\caption{Schematic view of the electric double layer with screening length $%
\protect\kappa ^{-1}$. The ions within the demixing volume of thickness $%
\protect\sigma $ diffuse toward higher water content. As a result, the
particle moves backward (cap at the rear) for hydrophilic coating, and
forward (cap at the front) for hydrophobic coating. }
\end{figure}

\textit{Charge effects.} Surface charges have been shown to be relevant for
the reversible aggregation of polystyrene particles in a near-critical
water-lutidine mixture \cite{Gal92}, and they may even change the sign of
the crtical Casimir effect \cite{Gam09}. Here we complete the above
discussion of self-diffusiophoresis by including charge effects, as a
possible explanation for the backward motion of beads with hydrophilic cap.

A composition gradient $\nabla\phi$ along a charged surface gives rise to
two distinct effects: the drift of the mobile counterions due to the
ion-specific thermodynamic force $-\nabla \mu$, and the non-uniform
properties of the electric-double layer, very much like the thermal forces
in a temperature gradient \cite{Wue10}. Here we discuss the ion-drift term
only, and reduce the chemical potential to the electrostatic self-energy of
a monovalent ion of radius $a_{m}$, 
\begin{equation}
\mu =\frac{e^{2}}{8\pi \varepsilon a_{m}}.  \label{32}
\end{equation}%
The variation of the composition $\phi $ in the demixing volume is
illustrated in Fig. 5. The dependence of the permittivity $\varepsilon $ on $%
\phi$ gives rise to a thermodynamic force density $-\rho \partial _{x}\mu $
with the ion concentration $\rho $. This ion current drags the fluid along
the particle surface and thus induces a slip velocity 
\begin{equation}
v_{s}=-\frac{1}{\eta }\int_{0}^{\infty }dzz\rho \partial _{x}\mu .
\label{34}
\end{equation}%
Spelling out the gradient, $\partial _{x}\mu =-\mu (\partial _{\phi }\ln
\varepsilon )\partial _{x}\phi $, assuming the linear law $\varepsilon =\phi
\varepsilon _{w}+(1-\phi )\varepsilon _{l}$ for the permittivity of
water-lutidine, and using $\varepsilon _{w}\gg \varepsilon _{l}$, one finds $%
\partial _{x}\mu= -\mu \partial _{x}\phi $.

The particle velocity $v_p$ is given by the surface average of the negative
slip velocity. Evaluating the counterion concentration in Debye-H\"{u}ckel
approximation, $\rho =(\varepsilon /e)|\zeta |e^{-\kappa z}$, we find \cite%
{supplmat} 
\begin{equation}
v_{p}=-\frac{e|\zeta |}{8\pi \eta a_{m}a}\int_{-1}^{1}dc\frac{%
(1-c^{2})\partial _{c}\phi }{\left[ 1+(\sigma \kappa )^{-1}\right] ^{2}},
\label{36}
\end{equation}%
which strongly depends on the ratio of the screening length $\kappa ^{-1}$
and the thickness $\sigma $ of the demixing volume. With typical parameters
the prefactor takes a value of millimeters per second. Taking $%
\sigma\kappa\sim\frac{1}{10}$ and $\partial_c\phi\sim0.1$ as suggested by
Figs. 1 and 2, we find $v_p\sim\mu$m/s, which corresponds to measured
values. So far we have considered the salt-free case where $\kappa^{-1}\sim$
hundreds of nanometers. Adding salt would result in diffusiophoresis in a
non-uniform electrolyte \cite{Ebe88,Abe08,Esl14}. Moreover, we have
discarded specific-ion effects which are not small in general \cite%
{Bro14,Wan14}.

According to (\ref{36}), charged particles move in the direction opposite to
the composition gradient, that is, cap at the rear for water-adsorbing
(hydrophilic) coating. This is precisely what was observed in experiments on
particles with hydrophilic gold caps \cite{Vol11,But12}. The ionic endgroups
used in these studies (11-mercapto-undecanoic-acid) cause a $\zeta$%
-potential of about $-50$ mV \cite{Laa06}, which results in a negative
velocity $v_p$ of microns per second.

On the other hand, the cap-at-the-front orientation, expected for
lutidine-adsorbing (hydrophobic) charged surfaces, is probably of little
relevance: The carbon caps \cite{But13} and hydrophobic gold caps gold caps
(functionalized with 1-octadecanethiol) \cite{But12} carry only weak
charges; as a consequence, their forward motion is due to the dispersion
forces underlying (\ref{30})

\textit{Conclusion.} Hot Janus particles in a near-critical water-lutidine
mixture move due to their self-generated composition gradient. Our analysis
reveals two main mechanisms: The dispersion forces exerted on water and
lutidine result in a positive velocity $v_p>0$, whereas at a charged
surface, the counterions migrate toward higher water content and thus drive
hydrophilic particles backward, $v_p<0$. The first effect accounts for the
observed forward motion of uncharged particles \cite{But12,But13}, and the
second one for the backward motion of beads with charged hydrophilic caps 
\cite{Vol11,But12}.

Both the velocity $v_p$ and the squirmer parameter $\beta$ show a
non-monotonous variation with the wetting properties and surface
temperature; this implies that the hydrodynamic interactions related to $%
\beta$ strongly depend on the driving power. The phase behavior of squirmer
systems is very sensitive to the value of $\beta$ \cite{Ish08,Llo10,Zoe14};
in view of Fig. 4 it could be changed by tuning the heating.

As an outlook, our findings, both on the wetting properties and on charge
effects, could be relevant for other driving mechanisms of active particles.
Moreover, they suggest that the overlap of the critical droplets of nearby
Janus particles should give rise to a complex interaction pattern, which
could affect the observed aggregation behavior and result in a variety of
reversible ordered states, similar to those realized recently with Janus
particles in a homogenous near-critical water-lutidine mixture \cite{Yu14}.

Helpful discussions with Y.\ Amarouch\`{e}ne, C.\ Bechinger, and U.\ Delabre
are gratefully acknowledged. This work was supported by Agence Nationale de
la Recherche through contract ANR-13-IS04-0003,

\end{document}